\documentstyle[12pt,a4]{article}

\def\simge{\mathrel{%
   \rlap{\raise 0.511ex \hbox{$>$}}{\lower 0.511ex \hbox{$\sim$}}}}
\def\simle{\mathrel{
   \rlap{\raise 0.511ex \hbox{$<$}}{\lower 0.511ex \hbox{$\sim$}}}}

\def\be{\begin{equation}}
\def\ee{\end{equation}}
\def\bea{\begin{eqnarray}}
\def\eea{\end{eqnarray}}
\def\ba{\begin{array}} 
\def\ea{\end{array}}

\begin{document}
\begin{center}

{\ \ }
\\ \bigskip 
{\Large \bf  \noindent
EQUIVALENCE\,\, PRINCIPLE\,\, TESTS } \\
\vspace{7mm}
{\Large \bf  \noindent AND\,\, NEW\,\, LONG-RANGE \,\,FORCES}
\medskip
\vglue 1.6 true cm

{\large \bf  \noindent  P. FAYET}

\vspace{5mm}
{\noindent \it
Laboratoire de Physique Th\'eorique de
l'Ecole Normale Sup\'erieure\,\footnote{~UMR 8549, 
~Unit\'e Mixte du CNRS et de l'Ecole Normale 
Sup\'erieure.   \\
e-mail: \ fayet@physique.ens.fr}, \\
24 rue Lhomond, 75231 Paris CEDEX 05, ~France}

\vskip 1.4truecm

\noindent
{ \bf ABSTRACT}
\end{center}

\parskip 5pt

\baselineskip 15pt

\vspace{.1truecm}
\small
{
We discuss the possible existence of new long-range forces 
mediated by spin-1 or spin-0 
particles. They would add their effects 
to those of gravity, and could lead to {\it apparent\,} 
violations of the Equivalence Principle.
Informations on the (vector and axial) couplings 
of a new spin-1 $\,U\,$ boson
may be obtained from spontaneously broken gauge invariance. 
The charge associated with the vector coupling
may be expressed as a linear combination of $\,B\,$ and $\,L\,$.
\,If the new force has a finite range $\,\lambda\,$,
\,its intensity 
turns out to be proportional to $\,{1/\, (\lambda^2\ F^2)},\,$
$\,F\,$ being the extra $U(1)\,$ symmetry-breaking scale.

Quite surprisingly, particle physics experiments can provide 
constraints on such a force, even if the corresponding gauge coupling 
is extremely small ($\ll \ 10^{-19}\ !)$. 
An ``equivalence theorem'' shows that a very light spin-1 $\,U$ boson 
with non-vanishing axial couplings does not in general decouple even when its 
gauge coupling vanishes, but behaves as a quasimassless pseudoscalar. 
(This equivalence theorem is similar to the one of
supersymmetry/super\-gravity theories, according to which 
a very light spin-$\frac{3}{2}\,$ gravitino 
might get detectable as a quasi massless spin-$\frac{1}{2}\,$
goldstino, 
despite the extreme smallness of Newton's gravitational constant
$\,G_N$.)
~Searches for the radiative production of such $\,U\,$ bosons in 
$\,\psi$ and $\,\Upsilon\,$ decays 
restrict the extra $\,U(1)\,$ symmetry-breaking scale $\,F\,$ 
to be larger than the electroweak scale, 
providing constraints on the intensity of the corresponding new force.

\normalsize

\vskip .2truecm

\vfill
\noindent
Based on a talk given at the 33$^{\rm rd}$ COSPAR Assembly 
``Fundamental Physics in Space'' (Warsaw, Poland), \,July 2000.
\begin{flushright}
\vskip -.3truecm
LPTENS--01/25   
\vspace*{-5mm}
\end{flushright}

\vspace{.3truecm}

\parskip 5pt
\baselineskip 15pt

\newpage

\section{\large \bf GENERAL OVERVIEW.}
\vspace{-.1truecm}

Why to test the Equivalence Principle\,? Could new long-range forces exist, 
in addition to gravitational and electromagnetic ones, and what could be 
their properties\,?

 The Equivalence Principle is at the basis of the theory of General Relativity. 
Although we have no reason to believe that general relativity is incorrect, 
it is certainly not a satisfactory, complete theory. 
In particular there is a well-known 
clash between general relativity and quantum physics. More precisely, no
consistent quantum theory of gravity exists, although one hopes 
to progress towards a solution within the framework of superstring 
and membrane theories. While the problem may be 
ignored temporarily for gravitational 
interactions   
of particles at physically accessible energies, it becomes crucial 
at very high energies of the order of the Planck energy, 
$\,\simeq 10^{\,19}\,$ GeV. \,This is the energy scale 
(corresponding in quantum physics to extremely small distances 
$\,\sim L_{\hbox{\footnotesize{Planck}}}\simeq 1.6\ \,10^{-33}\,$ cm) 
\,at which gravity is normally expected to become a strong interaction, 
so that quantum effects, still ill-defined, become essential. 
At such energies gravity has an effective intensity comparable to that 
of the three other interactions, strong, electromagnetic and weak. 
This is where a unification of all four interactions might 
conceivably occur.

Independently of gravity, the Standard Model of strong, electromagnetic 
and weak interactions is very successful in describing the physics of 
elementary particles and their fundamental interactions. 
But it also suffers from certain difficulties, and leaves a number 
of questions unanswered. To mention a few:

\vspace{-1truemm}
-- it has about 20 arbitrary parameters, including the three gauge 
couplings of $\ SU(3) \times SU(2) \times U(1)\,$, \,two parameters 
$\,\mu^2\,$ and $\,\lambda\,$ ultimately fixing the $\,W\,$ and Higgs 
boson masses, and thirteen mass and mixing-angle parameters associated 
with the quark and lepton spectrum. 

\vspace{-1truemm}
-- it sheds no light on the origin of the various 
symmetries and of symmetry breaking, nor on the family problem 
(why three generations of quarks and leptons, ...).

\vspace{-1truemm}
-- in the presence of very large mass scales it suffers from the problem 
of the stability 
of the mass hierarchy:  
how can the $\,W\,$ mass 
 remain so small compared to the grand-unification or the Planck scales, 
in spite of radiative corrections which  would tend to make it of the same 
order as  $\,m_{\hbox{\tiny GUT}}\,$ or 
$\,m_{\,\hbox{\footnotesize {Planck}}}\,$?

\vspace{-1truemm}
-- another problem concerns the vacuum energy, when coupled to gravity:
unless it is zero or almost zero (i.e. really extremely small, 
when measured with the natural units of particle physics, even more 
in terms of Planck's units),
it tends to generate a much too large value of the cosmological constant 
$\,\Lambda\,$, exceeding by many orders of magnitude 
what is experimentally allowed 
($\,|\Lambda|\,< \,3 \ 10^{-56}\ {\rm cm}^{-2} \,\simeq \,
\,10^{-121}\ L_{\hbox{\footnotesize{Planck}}}^{\ \ -\,2}\,!\,$).

\vspace{-1truemm}
-- a delicate question concerns the symmetry or asymmetry between Matter 
and Antimatter. The $\,CP\,$ symmetry is almost a symmetry of all 
interactions, but it is violated by some weak interaction effects 
observed in kaon decays. 
Then it has no reason to be an exact symmetry of strong interactions, 
so that the neutron should acquire an electric dipole 
moment. 
Since no such moment has been found the corresponding amount of 
$\,CP$-violation (measured by the dimensionless parameter 
$\,\theta_{\hbox{\footnotesize{eff}}}\,$)
\,should be smaller than $\,\approx 10^{-9},$
\,already a very small number. 
A possible mechanism to understand this 
requires the existence of a new neutral, very light, spin-0 particle, 
the axion (Wilczek, 1978; Weinberg, 1978).

	Essentially all attempts to go beyond the Standard Model and try 
to bring a solution to the above problems involve the introduction of 
{\it \,new symmetries}, {\it \,new particles}, and therefore quite possibly
{\it \,new forces}. 

	Such a situation already occurred thirty years ago, when the problems 
associated with the non-renormalisability of known (charged-current) weak 
interactions led physicists to rely on the new gauge symmetry principle, 
and to postulate the existence of a new particle, the neutral gauge boson 
$\,Z$, in addition to the (then still hypothetical)
charged ones, $\,W^+$ and $\,W^-$. 
\,This finally led to what is known as the Standard Model. The $\,Z$ mass 
had to be of the same order as the $\,W$ mass, $\,Z$-exchanges being 
responsible for a new class of weak-interaction effects 
(through ``neutral currents'') that were subsequently discovered in 1973, 
ten years before the $\,W$'s and $\,Z$'s could be directly produced at CERN, 
in 1983. The corresponding new force has a very small range of about 
$\,2 \ 10^{-16}$ cm, as for charged-current weak interactions.
It is now conceivable (and even likely)  
that the solution to the problems associated 
with the quantization of gravity requires the existence of new particles
\,-- in addition to the usual massless spin-2 graviton --\, 
and therefore of new forces,
possibly long-ranged, appearing as additions or modifications 
to the known force of gravity.

	Irrespectively of gravitation, the grand-unification between 
electroweak and 
strong interactions would involve very heavy spin-1 
gauge bosons that could be responsible for proton decay. The supersymmetry 
between bosons and fermions requires the existence of new superpartners 
for all particles. These new particles \,-- together with the two Higgs 
doublets required for electroweak breaking within supersymmetry --\, 
have a crucial effect on the evolution of the weak, electromagnetic 
and strong gauge couplings, 
allowing them to converge, at a large value of the grand-unification 
energy scale of the order of $\,10^{16}\,$ GeV.
\,Supersymmetry is also closely related with gravitation, since a locally 
supersymmetric theory must be invariant under general coordinate 
transformations. And the lightest of the new superpartners 
predicted by supersymmetric theories, which all have an odd 
$\,R$-parity character 
(with $\,R$-parity equal to $\,(-1)^{2\,S}\ (-1)^{3\,B+L}\,$)
\,turns out be to an almost ideal candidate to constitute the non-baryonic 
Dark Matter that seems to be present in the Universe.

More ambitious theories involve extended supersymmetry, new compact space 
dimensions of various kinds, and extended objects like superstrings 
and membranes, aiming at a completely 
unified description of all interactions, including gravity. 
They involve many new particles, including in general new 
neutral spin-1 or spin-0 bosons appearing as lower-spin partners or 
companions of the spin-2 graviton, etc.. 
The exchanges of such new particles could lead to new forces 
adding their effects to those of gravity. They could manifest 
experimentally through (apparent) violations of the Equivalence Principle
\,-- according to which the gravitational and inertial masses 
may be identified --\,
since what seems to be, experimentally, the force of gravity, 
might in fact be the superposition of gravity itself
(acting proportionally to masses) with some other additional 
new force(s) having different properties.

In particular, spin-1 bosons hereafter called 
$\,U$-bosons could gauge extra $U(1)\,$ symmetries
(cf. Fayet, 1990), as will be discussed in more details below. 
$\,U$-exchanges would be responsible for a new force involving
the {\it vector\,} part in the $\,U\,$ current, and 
expected to act on ordinary matter in an {\it additive\,} way, 
proportionally to a linear combination of the numbers of protons 
and neutrons, $\,Z\,$ and $\,N$, \,as we shall see. 
~If the $\,U$ boson is massless or almost massless with
an extra $\,U(1)\,$ gauge coupling $\,g"$ extremely small, 
the new force would superpose its effects to those of gravitation, 
leading to {\it apparent violations of the Equivalence Principle},
since the numbers of neutrons and protons in an object 
are not exactly proportional to its mass.
Newton's $\,1/r^2\,$ law of gravitation could also appear to be
violated, if the new force has a finite range.

The spin-0 dilaton (or ``moduli'', etc.) fields originating from superstring 
scenarios may well (or even should) remain massless; 
they are then generally expected to lead to 
excessively large deviations from the Equivalence Principle. 
However these fields could have their vacuum expectation values attracted 
towards a point at which they would almost decouple from matter 
(Damour and Polyakov, 1994). 
Their residual interactions could then be detected through 
extremely small (apparent) violations of the Equivalence Principle,
possibly at a level estimated to be of the order of 
$\ 10^{-12}\,$ to $\,10^{-24}\,$. 

The Equivalence Principle has already 
been tested to a very good level of precision, of about
a few $\,10^{-12}\,$ at large distances 
(Roll et al., 1964; Adelberger et al., 1990; Su et al., 1994).
Lunar laser ranging data also indicate that the acceleration rates 
of the (Fe/Ni-cored) Earth and the (silicate-dominated)
Moon towards the Sun are practically equal, to a level of precision 
slightly better than $10^{-12}$ (cf. Williams et al., 1996;
M\"uller et al., 1997; M\"uller and Nordtvedt, 1998). 
Strictly-speaking the interpretation of this result, however, 
also involves the consideration of gravitational-binding energies, 
in addition to the different compositions of the Earth and the Moon.

The sensitivity of Equivalence Principle tests
could be further improved by monitoring 
the relative motion of two test masses
of different compositions,
circling around the Earth, in a drag-free satellite.
The MICROSCOPE experiment (``MicroSatellite 
\`a Compensation de tra\^{\i}n\'ee pour l'Observation du
Principe d'Equiva\-lence''), whose construction has just been decided by CNES, 
aims at testing the validity of this principle at a level of precision of 
$\,10^{-15}$ (Touboul, 2000). The STEP experiment 
(``Satellite Test of the Equivalence Principle'') is a more ambitious project 
which aims at a level of sensitivity that could reach 
$\,\sim 10^{-17}\,- \,10^{-18}$ (Blaser et al., 1996; Vitale, 2000),
a considerable improvement by five orders of magnitude or more
compared to the present situation.

The test masses, incidentally,
cannot be taken spherical, but only cylindrical.
A potential difficulty is the existence of residual interactions 
between the higher multipole moments of the test masses
and the gravity gradients induced by disturbing masses within the satellite, 
which could lead to an unwanted signal simulating 
a ``violation of the Equivalence Principle''.
To minimize these effects one can use test masses approaching ideal forms of 
``aspherical gravitational monopoles'',
which are homogeneous solid bodies for which {\it \,all\, higher 
multipole moments vanish identically,\,} despite the lack of spherical 
symmetry (Connes et al., 1997)\,!

Should deviations from the Equivalence Principle be observed, further 
informations relying on data from several differential accelerometers 
could allow one to distinguish between new spin-1 or spin-0 induced forces, 
adding their effects to those of gravity. In the first case the new force 
is generally expected to act on a linear combination of baryon and 
lepton numbers $\,B\,$ and $\,L\,$ 
(which coincides in practice with a combination 
of the numbers of protons and neutrons, $\ Z=L\,$ and $\ N=B\!-\!L\,$). 
For spin-0 exchanges the new force 
may be expected to act effectively on a linear combination of $\,B\,$ 
and $\,L\,$ with electromagnetic (and chromodynamics) energies. 
Should such a force be found, testing several pairs of bodies of 
different compositions could allow one to distinguish between the spin-1 
and spin-0 cases.

\vspace{.2truecm}	

\section{\sloppy \large  \bf GENERAL FEATURES OF A NEW \hbox{SPIN-1} 
INDUCED FORCE.}

\vspace{-.2truecm}

\subsection{Possible extra ${\,\boldmath U(1)}\,$ gauge symmetries.}

     For spin-1 particles we can rely on the general principle of 
gauge invariance to determine 
the possible couplings of a spin-1 $\,U$-boson, and the expected 
properties of the corresponding force, should it exist (Fayet, 1990). 
To do so we first identify the possible extra $\,U(1)$ symmetries 
of a Lagrangian density, which are potential candidates for being gauged. 
This turns out to depend  crucially on the number of Higgs doublets 
responsible for the electroweak breaking. 
In the Standard Model there is no other $\,U(1)$ symmetry 
than those associated with the conservations of baryon  
and lepton numbers ($B\,$ and $\,L_i\,$), and with the weak
hypercharge $\,Y\,$ generating the $\,U(1)$ subgroup of 
\hbox{$\,SU(2) \times U(1)\,$}. \,More generally, in any renormalizable 
theory with only one Higgs
doublet, any $\,U(1)\,$ symmetry generator $\,F$ must
act on quarks and leptons as a linear combination:
\begin{equation}
F \ \,=\,\   \alpha \,B \,+\, \beta _i \,L_i \,+ \,\gamma \,Y    \ \ .
\end{equation}

Supersymmetric theories, however, require two Higgs doublets. 
This leaves room for an additional $\,U(1)\,$ invariance, since we may 
now perform independent 
phase rotations on these two doublets. With two Higgs doublets separately 
responsible for up-quark masses
($h_2$), \,and down-quark and charged-lepton masses ($h_1$), 
\,we now get:
\begin{equation}
\label{symgen}
F \ \, =\ \, \alpha \,B \,+\, \beta _i \,L_i \,+\, \gamma \,Y \ \,
+\,\  \mu \,F_{ax}    \ \  ,
\end{equation}
$F_{ax}\,$ being an extra $\,U(1)\,$ generator 
corresponding to a symmetry group $\,U(1)_A\,$ acting {\it \,axially\,} on 
quarks and leptons. 
This $\,U(1)_A\,$ itself or, more generally, an extra $U(1)\,$ generated by 
a linear combination as in (\ref{symgen}) was gauged, 
in the first supersymmetric models of 1976-1977, 
to trigger spontaneous supersymmetry breaking
without having to resort to soft supersymmetry-breaking terms\,\footnote{This, 
however, also raised the delicate question 
of anomaly cancellation, although anomalous $\,U(1)$'s 
might possibly be tolerated after all ...}.
Such models provided, very early, 
a natural framework for a possible new 
long-or-intermediate-range  ``fifth  force'' (Fayet, 1980, 1981, 1986a, 1986b),
which may now be considered, independently of supersymmetry
(for which, in any case, other methods of supersymmetry breaking 
are now generally employed).
Furthermore, in grand-unified  theories  with  large gauge groups including 
$\,SU(5)\,$ or $\,O(10)\,$, quarks are related to leptons, so that 
$\,B\,$ and $\,L\,$ no longer appear separately, but only through their
difference $\ B\!-\!L\,$. 
The general form of an extra $\,U(1)\,$ symmetry generator 
that could be gauged is then given by:
\begin{equation}
F \ \ =\ \  \eta \ \left(\ \frac{5}{2} \  (B-L)\, - \,Y \ \right) \ \,
+\, \ \mu \ F_{ax} \ \ \ .
\end{equation}

\subsection{Expression of the new ``charge'' $\,{\boldmath Q_5}\,$.}

\vspace{-.1truecm}
	To know on which quantity the new force should really act
(still within the framework of a renormalizable theory), 
we also have to take into account mixing effects between neutral
gauge bosons. The resulting $\,U$ current involves a linear combination 
of the extra-$U(1)$ current identified previously, 
with the $\,Z\,$ weak neutral current $\,J_Z = J_3 - \sin^2 \theta \ J_{em}\ $.
\,For simple Higgs systems the extra-$U(1)$ generator, and subsequently 
the $\,U$-current, does not depend on the quark generation considered. 
The new force should then act on quarks in a flavor-conserving and
generation-independent way. 
There should be no couplings to strangeness,
charm or beauty, nor on mass itself either 
(in which case no ``deviation from the Equivalence Principle'' 
would have to be expected).  Couplings to a linear combination 
of $\,B\,$ and $\,L\,$ with the electrical charge $\,Q$, as well as couplings 
involving particle spins, are expected instead  (Fayet 1986a, 1986b, 1990). 
They will originate from the {\it \,vector\,}  and {\it \,axial\,} parts in 
the $\, U$ current, respectively.

     The vector part in the $\,U$ current is found  to  be  a  linear
combination of baryonic and leptonic currents with  the  electromagnetic
current, associated with the (normally conserved) charge 
\begin{equation}
Q_5 \ =\  x \,B \,+ \,y_i \,L_i \,+ \,z\, Q_{el}        \ \ , 
\end{equation}
which reduces to 
\begin{equation}
Q_5 \ =\  x \,(B - L )\, +\, z\, Q_{el}  \ \ ,         
\end{equation}
in the framework of grand-unification. 
Even if the new force acts in general on electrons as well as 
on protons and neutrons, 
the above formulas further simplify, for ordinary neutral matter, into  
\begin{equation}
Q_5 \ =\  x \,B \,+\, y\, L \ =\  x \,(N+Z)\, +\, y \,Z  \ \ ,
\end{equation}
or, in grand-unification, to 
\begin{equation}
Q_5 \ \, = \,\  x \ (B-L) \ \,=\,\  x \ N  \ \ \ .
\end{equation}

     The action of such a \,\underline{spin-1-induced}\, fifth force  on
neutral matter may then be written in an \underline {\it additive\,} way, 
proportionally to a linear combination of the numbers of protons 
(and electrons) and neutrons, $\,Z\,$ and $\,N\,$. 
This is of course not true for the gravitational force itself, 
and has no reason to be true in the case of a force induced by 
spin-0 exchanges.
\,(As an illustration, this means, for example, 
that the $\,U$-induced force acting on a helium-4 atom 
should be twice the force acting on a deuterium atom
\,--\,
while the mass of this helium-4
atom differs from twice the mass of a deuterium atom.)\,
In the framework of grand-unification, 
$\,B\,$ and $\,L\,$ only appear through their difference so that 
the new force is expected to act effectively on neutrons only. 
When the $\,U\,$ boson is massless or almost massless and 
the extra $\,U(1)\,$ gauge coupling is extremely small, 
one expects (very small) violations of the  Equivalence
Principle, since the numbers of neutrons and protons in an object 
are not exactly proportional to its mass.

      Should such a force be discovered, its properties may be used 
to test its  origin, and whether it is due to spin-1 or spin-0 particles, 
for example, since a \,\underline{spin-0 induced}\, force has no reason 
to act additively, precisely on a linear combination of 
$\,B\,$ and $\,L\,$ (not to mention the very specific combination $\,B\!-\!L\,$
which could appear in grand-unified theories).

\subsection{A relation between range and intensity\,?}

\vspace{-.1truecm}
The next things one would like to know are, of course,
the possible {\it range} of the new force and its expected {\it intensity}, 
relatively to gravity.

The range could be infinite if the $U$ boson stays massless. This  occurs 
for example if there is only a single Higgs doublet 
(and no other  Higgses), so that the
$\ SU(3) \times SU(2) \times U(1) \,\times \, \hbox {extra-}U(1)\ $ 
gauge group gets broken down 
to $SU(3) \times  U(1)_{QED} \times   
U(1)_{\small U\hbox{-}\rm boson}\ $, the $\,U\,$ 
boson remaining exactly massless and coupled to a linear combination of 
the conserved $\,B\,$, $\,L\,$ and electromagnetic currents
(Fayet, 1989). The intensity 
of the new force, determined by the value of the extra $U(1)$ gauge coupling 
constant $g"$, \,remains, at this stage, essentially arbitrary. It might be 
extremely small, especially if the extra $U(1)$ turns out to be linked 
in some way with gravitational interactions.

In general, however, both the range of the new force and its possible 
intensity appear as largely arbitrary. Still {\it \,for any given symmetry 
breaking scale} $\,F\,$ these two quantities turn out to be related
as follows:
\vspace{-.05truecm}
\begin{center}
``\,\underline{\it the longer the range $\,\lambda$, \,the smaller the expected intensity}\,''.
\end{center}

\noindent
The origin of this interesting relation is in fact not mysterious.
The $\,U\,$ boson mass \,-- when it does not vanish --\,
determines the range of the corresponding force by the usual formula
of quantum physics:
\begin{equation}
\lambda \ =\ \frac{\hbar}{m_U\,c} \ \ 
\simeq \ \  2 \ \,\hbox {meters}\ \ \frac{10^{-7} \ \hbox{eV}/c^2 }{ m_U} 
\ \ \ .
\end{equation}
Large masses $\ \simge\,  200 \ \hbox{GeV}/c^2\ $ -- 
well within the domain of particle 
physics --\, would correspond to extremely short ranges 
$\,\lambda \simle \,10^{-16}\,$ cm, 
\,less than the range of weak interactions. 
On the other hand very small masses of $\ \, 10^{-10} \ \hbox{eV}/c^2\,$ 
or less, for example, 
would lead to 
large macroscopic ranges of \,2\, kilometers or more. The new force would 
then superpose its effects to those of gravitation, leading to apparent 
violations of the Equivalence Principle; and also, depending on the range, 
to apparent deviations from Newton's $\,1/r^2\,$ law of gravitation.

\vspace{6mm}
Let us now turn to the {\it \,intensity\,} of the new force, 
considered relatively to gravity. It may be 
characterized, at distances larger than the range $\,\lambda$, \,by
the dimensionless ratio
\begin{equation}
\tilde \alpha \ \ \approx \ \ 
\frac{\left(\hbox{\,\large $\frac{g"}{4}$\,}\right)^2 / 
\,4\pi} {G_{\hbox{\footnotesize Newton}}\  
m_{\hbox{\footnotesize proton}}^{\ 2}} \ \  
\approx \ \ 10^{36} \ \, g"^2   \ \ \ ,
\end{equation}
$g"\,$ being the extra-$U(1)$ gauge coupling constant, a priori unknown 
but which may be extremely small 
(especially, again, if the extra $\,U(1)\,$ symmetry 
turns out to be linked in some way with gravity itself).
The $\,U\,$ mass is related to  the
extra-$U(1)$ symmetry-breaking scale $\,F,\,$ determined by the appropriate 
Higgs v.e.v.'s, \,by a relation which may be written as
\begin{equation}
m_U \ \,\approx \ \,g"\ \,\frac{F}{2}  \ \ \ .
\end{equation}
For a given scale $\,F$, \,the  relative intensity of
the  new  force  then behaves like 
\begin{equation}
\tilde \alpha \ \ \sim \  \  g"^2  \ \ \sim \ \ 
\frac{{\,m_U}^2\,}{F^2}\ \ \sim \ \ \frac{1}{\lambda^2 \ \  F^2}\ \ \ ,       
\end{equation}
or, more precisely, with an uncertainty 
reflecting the effects of the model-dependent factors in the coefficients
(Fayet, 1986a, 1986b):
\begin{equation}
\label{alpha}
\tilde \alpha \ \ \ \approx \ \ \ 
\frac{1}{\lambda\,(\hbox{\footnotesize {meter}})^2} \ 
\left( \  \frac{250\ \,\hbox{GeV}}{F}\  \right)^2    \ \ \ .  
\end{equation}

\vskip .2truecm

This relation looks very nice, but before we can really use it
we must know or assume something 
about the symmetry-breaking scale $\,F\,$.
Before discussing this point in the next sections, 
let us assume for the moment, as an illustration, that 
the extra $\,U(1)\,$ is broken at or around the electroweak scale
$(\,\approx \ 250 \ \,\hbox{GeV}\,)$, a natural benchmark in particle physics. 
We would then get for small (or moderate) values of the range $\,\lambda$, 
\,rather large (or not so small) values of $\,\tilde\alpha$,\, e.g.
\begin{equation}
\tilde \alpha  \ \ \ \approx \ \ \left\{  \begin{array}{ccl}
\,10^{7} \,- \,   10^{9}\,   &\ \ \ \ \hbox {if} \ \ \ &\lambda \ 
\simeq \ 10^{-1} \ \hbox{mm} \ \ , \vspace{2.5mm}\cr
10^{-3} - 10^{-5}  & \ \ \ \ \hbox {if} \ \ \ &\lambda \ 
\simeq \ \ 100 \ \hbox{m}  \ \ \ ,  \cr   \end{array}   \right.  
\end{equation}
values which are already forbidden by existing gravity 
experiments including those performed 
at short distances (Hoskins et al., 1985; Mitrofanov and Ponomareva, 1988;
Adelberger et al., 1990; Su et al., 1994; Lamoreaux, 1997;
Schmidt et al., 2000; Hoyle et al., 2001),
which imply, for example, that $\,\tilde \alpha\,$ should be smaller 
than $\,10^{-1}\,$ for ranges $\,\lambda\, \simge \, 1$ mm).
This corresponds to the fact that in such cases 
the new extra $\,U(1)\,$ gauge coupling $\,g"\,$ is not so small compared to 
$\,10^{-19}\,$ or even significantly larger 
in the case of a small $\,\lambda,$ 
\,so that the new force is not so small compared to gravity
or may even dominate it at small distances.

\vspace{3mm}

On the other hand we would have
\begin{equation}
\label{alphatilde}
\tilde \alpha  \ \ \ \approx \ \ \left\{  \begin{array}{ccl}
\,10^{-5} \,- \,  10^{-7}\,   &\ \ \ \ \hbox {if} \ \ \ &\lambda \ 
\simeq \ \ \ 1 \ \, \hbox{km} \ \ , \vspace{2.5mm}\cr
10^{-11} - 10^{-13}  & \ \ \ \ \hbox {if} \ \ \ &\lambda \ 
\simeq \ 10^3 \ \hbox{km}  \ \ ,  \cr   \end{array}   \right.  
\end{equation}
the latter case, which would lead to apparent violations of the
Equivalence Principle at the level of $\,10^{-13}\,$ to $\,10^{-16}\,$, 
\,being within the reach of the future MICROSCOPE and STEP experiments
\,-- sensitive only to ranges $\,\lambda \,$ larger than 
a few hundreds of kilometers, given the elevation at which the satellite 
should orbitate.
But, as we have already indicated, these estimates for $\,\tilde \alpha\,$ 
depend crucially on what the 
extra-$U(1)\,$ symmetry breaking scale $\,F\,$ is. 
Could there be some way to learn something about it\,? 

\vspace{3mm}
No one would imagine, under normal circumstances, 
being able to search directly for ordinary massless
{\it gravitons\,} in a particle physics experiment, 
due to the extremely small value of 
the Newton constant
($10^{-38}$, in units of \,GeV$^{-2}$), which determines the strength 
of the couplings of a single massless graviton to matter. 
Then how could we search directly,
in particle decay experiments, for $U$-bosons
with even smaller values of the corresponding coupling, 
$\ g"^2\ll 10^{-38}~$? Still this turns out to be possible\,!
This rather astonishing result holds as soon as 
the $U$-current includes a 
(non-conserved, as the result of spontaneous symmetry breaking) 
{\it axial\,} part, as we shall see. 
The origin of this phenomenon involves an ``equivalence theorem'' 
between the interactions of spin-1 gauge particles 
and those of spin-0 particles, 
in the limit of very small gauge couplings.

\vspace*{.2truecm}

\section{\large \bf \sloppy
``EQUIVALENCE THEOREMS'' FOR SPIN-1 AND SPIN-$\frac{3}{2}\,$ PARTICLES.}

\vspace{-.2truecm}

\subsection{A very light spin-1 
$\,{\boldmath U}\,$ boson does ${\boldmath not\,}$ decouple for vanishing 
gauge coupling -- but behaves like a spin-0 particle\,!} 

     One might think that, in the limit of vanishing extra-$U(1)\,$ 
gauge coupling constant $\,g"$, \,the effects of the  new  gauge  boson 
would be arbitrarily small, and may therefore be disregarded (as for 
graviton effects in particle physics). But in general this is {\it wrong\,}, 
as soon as the $\,U$-current involves a (non-conserved) axial part 
\,-- which is generally the case when a second Higgs doublet is present 
to break the electroweak symmetry, as in 
supersymmetric theories\,!

     The amplitudes for emitting a very light ultrarelativistic $\,U\,$
boson are proportional to the new gauge coupling $\,g"$, \,and therefore
seem to vanish with $\,g"$. 
This is, however, misleading, since the polarization vector for
a longitudinal $\,U\,$ boson of four-momentum $\,k^\mu$,
$\,\epsilon^{\mu} \simeq \,k^{\mu}/m_U\,$, 
becomes singular in this limit, since $\,m_U \approx g"\, F/2\,$  
also vanishes with $\,g"$. \,Altogether the amplitudes for emitting, 
or absorbing, 
a longitudinal $\,U\,$ boson, appear to be essentially proportional 
to $\,g"/m_U\,$.
They have a finite limit,
{\it \,independent of} $\,g"$, \,when this gauge coupling becomes very small 
and the mass of the $\,U\,$ boson gets also very small, so that this $\,U\,$ 
boson is ultrarelativistic (i.e. $\,k^\mu \gg m_U$). 
\,Such a $\,U\,$ boson then behaves very much like a spin-0 particle 
(Fayet 1980, 1981), somewhat reminiscent of an axion.    

    This ``equivalence theorem'' expresses that in the high-energy or 
low-mass limit ($\,E \gg m_U\,$), the third (longitudinal) degree of 
freedom of a massive $U$-boson continues to behave like the 
massless Goldstone boson which was ``eaten away''. For very small $\,g"$ 
 the  \hbox{spin-1} $\,U$-boson simply behaves as this massless spin-0 
Goldstone boson. 
This applies as well to virtual exchanges. The exchanges of the $\,U$ 
boson do not disappear in this limit, owing to the non-conserved axial part 
in the $\,U$ current 
(in general present when there is more than one Higgs doublet). 
They become equivalent to the exchanges 
of a  massless  (pseudoscalar, $\,CP$-odd) spin-0  particle $\,a$, 
having effective axionlike couplings to leptons and  quarks
\begin{equation}
\label{coupl0}
 2^{1/4} \ \,G_F^{\ 1/2} \ \  
m_{\,l,q} \ \ \,(x \ \,\hbox {or} \,\ 1/x) \ \ \ \times\ 
\left(\ r\, \approx\,\frac{\, 250 \ \hbox{GeV}\,}{F}\ \right)  \ \ \gamma_5 \ \ ,
\end{equation}  
$x$ denoting the ratio of the two Higgs doublet  vacuum  expectation
values. 
Using notations which are standard in supersymmetry\,\footnote{The
pseudoscalar $\,a$, here eaten away to become the third (longitudinal) 
degree of freedon for the massive $\,U\,$ boson, 
is of course by construction reminiscent 
of the well-known pseudoscalar $\,A\,$ of supersymmetric extensions 
of the standard model, which becomes the Goldstone boson 
of the $\,U(1)\,_A\,$ invariance when this one is taken as a
symmetry of the Lagrangian density.}, 
where the first Higgs doublet is responsible for down-quark 
and charged-lepton masses, and the second one for up-quark masses,
one has 
$1/x = v_2/v_1 $ $= \tan \beta\,$.
\ More precisely, these effective pseudoscalar couplings to quarks and leptons 
read

\be
\label{coupl}
\hbox{\small $
\underbrace{
\underbrace{2^{1/4} \ \,G_F^{\ 1/2} \ \ m_{\,l,q} \ \ \, 
\left\{ 
\ba{cc}
x \ \hbox{(i.e. $1/\tan\beta$}) 
&\hbox{for \ \ \ \,$u,\ c,\ t\ $ \ quarks} 
\vspace{2mm} \\
1/x  \ \ \hbox{(i.e. $\tan\beta$})
&\ \ \hbox{for} \   
\left\{ \!\!\!
\hbox{
\begin{tabular}{c}
$d,\ s,\ b\ $     \  quarks    \vspace{1mm}  \\
$e,\ \mu,\ \tau\ $   leptons
\end{tabular}
} \!\!\!\!\!
\right.
\ea    
\right.
\!\!\!\!}_
{\hbox{\phantom\underline{\phantom\underline{\phantom\underline{
as for a standard axion}}}}} 
\ \times\ 
\left(\ r\, \approx\,\frac{\, 250 \ \hbox{GeV}\,}{F}\ \right)
}_
{\hbox{as for an ``invisible'' axion, \ \ \ \ if $\ r \ll 1$}}\ ,
$}
\ee

From there one can get, from particle physics, constraints 
on such a new spin-1 gauge boson, as we shall discuss 
in section \ref{sec:implications}. They will require 
the extra $U(1)$ symmetry-breaking scale $\,F\,$ to be larger than the 
electroweak scale. Note that the {\it tranverse\,} polarization states 
of the $\,U$-boson still continue to behave as usual, and would be 
responsible, for small but non-vanishing $\,g"$, \,for a very weak 
long-ranged ``EP-violating'' force, of intensity proportional to $\,g"^2$, 
\,to which we shall return in section \ref{sec:implications}.

\subsection{\sloppy Increasing the sym\-me\-try-breaking scale $F$ 
(``invisible \ \ ${\boldmath U}$-boson'' and ``invisible axion'' mechanisms).} 

\vspace{-.1truecm}
If the extra $\,U(1)\,$ gauge symmetry is broken at the electroweak scale 
\linebreak ($\,F \approx 250\ \, \hbox{GeV}\,$) by the two Higgs doublets 
$h_1$ and $h_2$ only,
the spin-1 $\,U$-boson acquires, from its non-vanishing axial couplings, 
exactly the same 
effective pseudoscalar couplings (\ref{coupl}) as a ``standard'' spin-0 axion
(i.e., $r \equiv 1$). Just as the latter, it then 
turns out to be excluded by the results of $\,\psi\,$ and $\,\Upsilon\,$ 
decay experiments, i.e. searches for the decays 
$\,\psi\ \to \,\gamma \,+\, \hbox{``nothing''}$, 
$\,\Upsilon\ \to \,\gamma \,+\, \hbox{``nothing''}$, \,in which ``nothing''
stands for a quasimassless neutral spin-1 particle (the $U$-boson), 
or a spin-0 particle (such as the axion), 
remaining undetected in the experiments.
{\it Exit\,} such a $U$ boson, and therefore the 
corresponding new force  that could be due to $\,U$-boson exchanges\,?

Not necessarily\,! As we observed in 1980, 
to save the possibility of such a light $\,U\,$ boson
(or just as well, to save the idea of the axion),
one can introduce an extra Higgs 
singlet acquiring a large v.e.v. $\gg\,250 $ GeV, 
which would make the $\,U\,$ boson significantly heavier 
(but still remaining light!)
without modifying the values of its (vector and axial) couplings 
to ordinary quarks and leptons.
Its effective interactions with quarks and leptons,
fixed by the ratio of the axial couplings 
\,-- proportional to $\,g"$ --\,
to the mass $\,m_U$, may then become arbitrarily small,
as one sees easily from the expression (\ref{coupl}) 
of the resulting effective peudoscalar couplings to quarks and leptons.
The mechanism, which involves for the effective interaction 
of the $\,U\,$ boson, or of its equivalent pseudoscalar $\,a$, 
the suppression factor
$\,r \approx \,F/250\ \hbox{GeV}\,\ll 1\,, $
\,allows for making the $\,U$ boson effects in particle
physics practically ``invisible'', provided the  extra $\,U(1)\,$ is broken 
``at a large scale'' $\,F\,$ significantly higher than  the  electroweak
scale.

Incidentally since a spin-1 $U$-boson, when very light, is produced and 
interacts very much as a spin-0 pseudoscalar axionlike particle
(excepted that it does not decay into two photons), the mechanism 
we explained also provided us, at the same time, with a way 
to make the interactions of the axion almost ``invisible'', at least in 
particle physics (Fayet 1980, 1981). This can be realized by breaking the 
corresponding global $\,U(1)_A\,$ symmetry, 
then considered as a Peccei-Quinn symmetry, at a very large scale, 
through a very large singlet vacuum  expectation value, the resulting axion 
being mostly an electroweak singlet. 
We thus obtained simultaneously both the  ``invisible $\,U$-boson'' mechanism
(for a spontaneously-broken extra $U(1)\,$ local gauge symmetry, 
the case of interest to us here), 
and the ``invisible axion'' mechanism (in the case of 
a global $U(1)_{\hbox{\tiny{PQ}}}\,$  symmetry broken at a very high scale) 
that became popular later
(Fayet 1980, 1981; Zhitnisky, 1980; Dine et al. 1981).

\subsection{The gravitino/goldstino ``equivalence theorem'' 
in supersymmetric theories.}
\vspace{-.1truecm}

The same phenomenon, in the case of local supersymmetry,
called supergravity, expresses that a {\it very light\,} spin-$\frac{3}{2}$ 
{\it \,gravitino} (the superpartner of the spin-2 graviton, 
and also the gauge particle of the local supersymmetry), 
having interactions fixed by the gravitational ``gauge'' coupling constant
$\,\kappa = \sqrt{\,8\,\pi\ G_N}\,\simeq\, 4.1\ \,10^{- 19}\,$ (GeV)$^{-1}$,
~would behave very much like a massless spin-$\frac{1}{2}\,$ 
{\it goldstino}, according to the ``equivalence theorem'' of supersymmetry 
(Fayet, 1977, 1979).
~Just as the mass of the $\,U\,$ boson is given in terms of the 
extra $\,U(1)\,$ gauge coupling $\,g"\,$ and symmetry breaking scale $\,F\,$ 
by the formula
$\ m_U\ \approx\ g"\,F/2$,
~the mass of the spin-$\frac{3}{2}$ gravitino is fixed by its 
(known) gravitational ``gauge'' coupling constant $\,\kappa\,$ and 
the (unknown) supersymmetry-breaking scale parameter $\,d$, ~as follows:
\be
m_{\hbox{\tiny{3/2}}}\ \ =\ \ \frac{\kappa\ d}{\sqrt 6}\ \ \simeq \ \ 1.68\ 
\left(\ \frac{\sqrt d}{100\ \hbox{GeV}}\right)^2\ 10^{-6}\ \hbox{eV}/c^2\ \ .
\ee
\noindent
The interactions of a light gravitino are in fact determined by the ratio
$\,\kappa/m_{\hbox{\tiny{3/2}}}\,$, ~or
$\ G_N/ m_{\hbox{\tiny{3/2}}}^{\ 2}\,$. ~As a result a sufficiently light 
gravitino might be detectable in particle physics experiments, 
despite the extremely small value of the Newton constant $\,G_N\,\simeq\,
10^{- 38}\,$ (GeV)$^{-2}$,
~provided the supersymmetry-breaking scale\,\,\footnote{An equivalent notation 
makes use of a parameter $\,\sqrt F\,=\,\sqrt d\,/2^{\hbox{\tiny{1/4}}}$,
~defined so that $\,F^2 = d^2/2$.
~Furthermore, the supersymmetry-breaking scale ($\,\sqrt d\,$ or $\,\sqrt F\,$) 
associated with a (stable or quasistable) light gravitino should in principle 
be smaller than a few $\,10^6$ GeV\,'s, 
~for its mass to be sufficiently small
($\,m_{\hbox{\tiny{3/2}}}\,\simle \,1 $ keV$/c^2$), so that relic gravitinos 
do not contribute too much to the energy density of the Universe.
} 
$\sqrt d\ $ is not too large.
The gravitino would then be the lightest supersymmetric particle,
with all other $\,R$-odd superpartners 
expected to ultimately produce a gravitino 
among their decay products, if $\,R$-parity is conserved.
(In particular the lightest neutralino could decay into photon + gravitino, 
~so that the pair-production of ``supersymmetric particles'' 
could lead to final states including two photons with missing energy
carried away by unobserved gravitinos.)

\vskip .2truecm

For a sufficiently light gravitino one can also search 
for the {\it direct production\,} of a
single gravitino associated with an unstable photino $\,\tilde \gamma$
~(or more generally a neutralino), decaying into
gravitino + $\,\gamma$, ~in $\,e^+e^-\,$ annihilations.
Or for the radiative pair-production of two gravitinos
in $\,e^+e^-\,$ or $\,p\,\bar p\,$ annihilations 
at high energies (Fayet, 1982, 1986c; Brignole et al., 1998a, 1998b),
e.g.
\be
e^+ e^- \ \ (\hbox{or}\ \ p\,\bar p)\ \ \ \rightarrow\ \ \ 
\gamma\ \ (\,\hbox{or jet\,)}\ \ +\ \ 
2\ \,\hbox {unobserved gravitinos}\ \ ,
\ee
which have cross-sections 
\be
\sigma \ \  \propto\ \ 
\frac{G_N^{\ 2} \ \ \alpha\, (\,\hbox{or} \ \,\alpha_s\,)\ \,s^3}
{m_{\hbox{\tiny{3/2}}}^{\ \ 4}}\ \ \propto\ \ 
\frac{\alpha\ (\,\hbox{or} \ \,\alpha_s\,)\ \,s^3}{d^4}\ \ \ .
\ee
Although the existence of so light gravitinos may appear 
as relatively unlikely, such experiments are sensitive to gravitinos of 
mass $\,m_{\hbox{\tiny{3/2}}} \simle\,10^{-5}\,$ eV/$c^2$, 
~corresponding to supersymme\-try-breaking scales 
smaller than a few hundreds of GeV's.

\vspace*{.2cm}

\section{\large \bf \sloppy IMPLICATIONS OF PARTICLE PHYSICS EXPERI\-MENTS.}
\label{sec:implications}

\vskip -.15truecm

Without necessarily having to consider very large values
of $F$, we can use formula (\ref{coupl}) (obtained with two Higgs doublets 
and an axial part in the $\,U$-current) to write  the branching ratios 
for the radiative production of $\,U\,$ bosons in quarkonium decays, 
proportionally to $\  r^2\,$ (or $\,1/F^2$):
\begin{equation}   \left\{   \  \begin{array}{ccll}
B \ (\ \psi \ \to \ \gamma \,+\, U\ ) &\simeq &
  \ 5 \ \ 10^{-5} \ \ \,\ r^2 \ x^2 \  \ \ &C_{\psi}  \nonumber 
\vspace{2mm}\cr
B\  (\ \Upsilon \ \to \ \gamma\, +\, U\ ) &\simeq &
 \  2 \ \ 10^{-4} \ \ \,(r^2/x^2) \ \  &C_{\Upsilon}            \cr
\end{array}   \right.
\end{equation}
($C_{\psi}$ and $C_{\Upsilon}$, expected to be larger than $1/2$, take 
into account QCD radiative and relativistic corrections). 
The $\,U\,$ boson, quasistable  or  decaying  into $\,\nu\, \bar \nu\,$,
 would remain undetected 
(as for an axion decaying into two photons outside the detector). 
\,From the experimental limits (Edwards et al., 1982; Crystal Ball coll., 1990; 
CLEO coll., 1995): 
\begin{equation}   \left\{   \  \begin{array}{ccl}
B\ (\ \psi \ \to \  \gamma \,+\,  \hbox{``nothing''}\ ) \ & <&\ \, 1.4 \ \ 10^{-5}\ \ ,  
\vspace{2mm}\cr
B\ (\ \Upsilon \ \to \ \gamma \,+\, \hbox{``nothing''}\ ) \  & < & \ \, 1.5 \ \ 10^{-5}
\ \ , \cr
\end{array}   \right. 
\end{equation}
we deduce 
$\ r  \simle 1/2\, $,
i.e. that the extra-$U(1)$ symmetry should be broken at a scale $\,F$ 
at least of the order of twice the electroweak scale (Fayet, 1980, 1981, 1986a, 1986b).

\vskip .3truecm 
This result, obtained for a $\,U$ with {\it non-vanishing axial couplings},
can be translated (assuming vector and axial parts in the $U$ current to be 
of similar magnitudes)  
into an approximate upper bound on the relative strength
of the new force, as a function of its range $\,\lambda\,$:
\begin{equation}
\label{const}
\tilde \alpha \ \approx
 \ \frac{\left(\hbox{\,\large $\frac{g"}{4}$\,}\right)^2 / 
\,4\pi} {G_{Newton}\  m_{proton}^{\ 2}}  \  
\approx \ \frac{1}{\lambda\,(\hbox{\footnotesize {meter}})^2} \ 
\left(\, \frac{250\ \hbox{GeV}}{ F}\, \right)^2 \ 
\simle\  \frac{1}{\lambda\,(\hbox {meter})^2 }  \ \ \ . \ \ 
\end{equation}

\vspace{.1truecm}
These particular constraints allow for a new force that, 
if it had a short range, could be very large 
compared to the gravitational force 
(e.g. up to $\,\tilde \alpha\, \approx \, 10^6\,$ for a range 
$\,\lambda \simeq 1$ millimeter, for example), but such large values 
are already forbidden by short-range gravity experiments (Hoyle et al., 2001).
If, on the other hand, the range  $\,\lambda\,$ turns out to be large,
the constraints (\ref{const}) become quite significant, and may be used 
to restrict or even practically exclude the existence of the new force 
of the special type considered here,
whose relative intensity is then experimentally constrained to be rather small.
As an illustrative example with $\,\lambda = 10^3$ km, we would get 
$\ \tilde  \alpha \,\simle \, 10^{-11} - 10^{-13}$, 
\,corresponding to expected violations of the Equivalence Principle 
\be
\simle  \ 10^{-13} - 10^{-16} \ \ ,
\ee
with an upper bound still within the sensitivity of the MICROSCOPE 
and STEP experiments. 
For significantly larger 
$\,\lambda\,$'s, however, the violations are likely to remain undetectable, 
in the present case of a spin-1 $U$-boson with non-vanishing axial couplings.

\section{\large \bf CONCLUSIONS.}
\vskip -.2truecm

We emphasize that the above constraints of section \ref{sec:implications}, 
which are rather drastic in the case of a long-range force,
concern the case of a $\,U$ boson having non-vanishing axial couplings,
on which we have been concentrating here.
For a $U$-boson coupled to a {\it purely vectorial\,} current, 
on the other hand \,--  or in the case of a new force due to spin-0 exchanges
--\, 
no such constraints are obtained.
Then the strength of the new force and its range (finite or infinite) remain
unrelated parameters.

\vspace{1mm}
We also mention, in addition, that the exchanges 
of a new spin-1 $\,U\,$ boson, or of a spin-0 particle 
such as the axion, could lead to new forces acting on particle spins.
More precisely one can search for a ($CP$-conserving) spin-spin interaction, 
and a $(CP$-violating) ``mass-spin coupling'' interaction 
(Moody and Wilczek, 1982; Fayet, 1996), 
which we do not discuss here.

\vspace{1mm}
Very precise tests of the Equivalence Principle could be sensitive
to new spin-0 or spin-1 induced forces, and could in principle allow 
us to distinguish between the two possibilities of spin-0 or spin-1 
induced forces. 
For a spin-1 $\,U$ boson with
 non-vanishing axial couplings the intensity of the new force is in general
constrained to be extremely weak if it is long-ranged, 
otherwise it remains essentially a free parameter. 
Testing, to a very high degree of precision, the Equivalence Principle 
in Space would bring new constraints on Fundamental Physics, and might 
conceivably lead to the spectacular discovery of a new long-ranged 
interaction, should a deviation from this Principle be found.

\vskip .8truecm

\noindent
{\large \bf REFERENCES}

\vskip .5truecm

\baselineskip6pt

\small

\baselineskip6pt
\noindent
Adelberger, E.G., et al., {\em Phys. Rev.} D {\bf 42}, 3267, 1990.

\noindent
Blaser, J.-P., et al., STEP,
Report on the Phase A Study, ESA report SCI ({\bf 96}) 5, 1996.

\noindent
Brignole, A., et al., 
{\em Nucl. Phys.} B {\bf 516}, 13, 1998a [erratum B {\bf 555}, 653, 1999].

\noindent
Brignole, A., et al., 
{\em Nucl. Phys.} B {\bf 526}, 136, 1998b [erratum B {\bf 582}, 759, 2000].

\noindent
CLEO Coll., {\em Phys. Rev.} D {\bf 51}, 2053, 1995.

\noindent
Connes, A., T. Damour, and P. Fayet, {\em Nucl. Phys.} B {\bf 490}, 391, 1997.

\noindent
Crystal Ball Coll., {\em Phys. Lett.} B {\bf 251}, 204, 1990.

\noindent
Damour, T., and A.M. Polyakov, {\em Nucl. Phys.} B {\bf 423}, 532, 1994.

\noindent
Dine, M., W. Fischler, and M. Srednicki, 
{\em Phys. Lett.} B {\bf 104}, 199, 1981.

\noindent
Edwards, C., et al., {\em Phys. Rev. Lett.} {\bf 48}, 903, 1982.

\noindent
Fayet, P., {\em Phys. Lett.} B {\bf 70}, 461, 1977.

\noindent
Fayet, P., {\em Phys. Lett.} B {\bf 86}, 272, 1979.

\noindent
Fayet, P., {\em Phys. Lett.} B {\bf 95}, 285, 1980.

\noindent
Fayet, P., {\em Nucl. Phys.} B {\bf 187}, 184, 1981.

\noindent
Fayet, P., {\em Phys. Lett.} B {\bf 117}, 460, 1982.

\noindent
Fayet, P., {\em Phys. Lett.} B {\bf 171}, 261, 1986a; B {\bf 172}, 363, 1986b;
B {\bf  175}, 471, 1986c.

\noindent
Fayet, P., {\em Phys. Lett.} B {\bf 227} 127, 1989.

\noindent
Fayet, P., {\em Nucl. Phys.} B {\bf 347}, 743, 1990.

\noindent
Fayet, P., {\em Class. Quant. Grav.} {\bf 13}, A19, 1996.

\noindent
Hoyle, C.D., et al., {\em Phys. Rev. Lett.} {\bf 86}, 1418, 2001.

\noindent
Hoskins, J.K., et al., {\em Phys. Rev.} D {\bf 32}, 3084, 1985

\noindent
Lamoreaux, S.K., {\em Phys. Rev. Lett.} {\bf 78}, 5, 1997.

\noindent
Long, J.C., H.W. Chan, and J.C. Price, {\em Nucl. Phys.} {\bf 39}, 23, 1999.

\noindent
Mitrofanov, V.P., and O.I. Ponomareva, {\em Zh.$\!$ Eksp.$\!$ Teor.$\!$ Fiz.}
{\bf 94}, 16, 1988 \, [{\em Sov.$\!$ Phys.$\!$ JETP\,} 
{\bf 67}, 1963, 1988].

\noindent
Moody, J.E., and F. Wilczek, {\em Phys. Rev.} D {\bf 30}, 130, 1984.

\noindent
M\"uller, J., et al., Proc. 8th Marcel Grossmann Meeting, Jerusalem,
1151, 1997.

\noindent
M\"uller, J., and K. Nordtvedt, {\em Phys. Rev.} D {\bf 58}, 062001, 1998.

\noindent
Roll, P.G., R. Krotkoov, and R.H. Dicke, {\em Ann. Phys.} {\bf 26}, 442, 1964.

\noindent
Smith, G.L.,  et al,, {\em Phys. Rev.} D {\bf 61}, 022001, 2000.

\noindent
Su, Y., et al., {\em Phys. Rev.} D {\bf 50}, 3614, 1994.

\noindent
Touboul, P., MICROSCOPE, in {\em Advances in Space Research}, this issue.

\noindent
Vitale, S., The STEP Project, in {\em Advances in Space Research}, this issue.

\noindent
Weinberg, S., {\em Phys. Rev. Lett.} {\bf 40}, 223, 1978.

\noindent
Wilczek, F., {\em Phys. Rev. Lett.} {\bf 40}, 279, 1978.

\noindent
Williams, J.G., X.X. Newhall, and J.O. Dickey, {\em Phys. Rev.} D {\bf 53} 
6370 (1996).

\noindent
Zhitnisky, A.P., {\em Yad. Fiz.} {\bf 31}, 497, 1980 \,
[{\em Sov. J. Nucl. Phys.} {\bf31}, 260, 1980].

\end{document}